\documentstyle[]{l-aa}

\begin{document}

  \thesaurus{05         % A&A Section 5: Stellar clusters and associations.
	     (02.03.1;  % Chaotic phenomena,
	      05.03.1;  % Celestial mechanics, stellar dynamics,
	      08.02.3;  % Stars: binaries: general,
              08.05.3;  % Stars: evolution, 
              08.12.3;  % Stars: luminosity function, mass function,
	      10.15.1)} % Galaxy: open clusters and associations: general.

\title{ The initial mass function and the dynamical evolution 
of open clusters}

\subtitle{IV. Realistic systems}

\author{R. de la Fuente Marcos} 

\institute{ 
           Universidad Complutense de Madrid, E--28040 Madrid, Spain 
           (fiast05@emducms1.sis.ucm.es) 
          }

\date{Received  ; accepted  }

\maketitle

%\maintitlerunninghead{IMF & open clusters IV} 
 
\begin{abstract} 
This paper is the fourth one of 
a series whose chief objective is studying the influence of different 
mass spectra on the dynamical evolution of open star clusters.
Results from several $N$-body calculations with primordial
binaries and mass loss due to stellar evolution are presented. The models  
show significant differences with those for primordial binaries but
no stellar mass loss presented in de la Fuente Marcos (1996b). A 
differential dynamical behaviour depending on cluster richness is 
found compared to de la Fuente Marcos (1996a). The evolution of 
these realistic models is very dependent on the initial mass function. 
Even for rich clusters, there is a dependence on the binary mass spectrum. 
The velocity distribution of the escapers is examined and
compared with results from previous calculations.
The evolution of the primordial binary population is   
analyzed in detail. The cluster remnant and the final binary population
are also studied. Finally, some conclusions about observational properties of 
Open Cluster Remnants are presented.


\end{abstract}

\section{Introduction}
In the last few years, a significant number of 
pre-main-sequence binaries and multiple systems have been 
discovered in young open clusters (Brandner et al. 1996; Ghez et al. 1993, 1994; 
Leinert et al. 1991, 1993; Padgett et al. 1996; Prosser et al. 1994;
Richichi et al. 1994; Simon et al. 1992, 1993, 1995). 
All the surveys carried out find a binary frequency which is greater than or
equal to that of the field stars in the range of separations to which
they are sensitive.
The observational data suggest that the binary cluster population
can be interpreted in terms of different formation mechanisms. Wide
binaries (with periods greater than 100 years) could originate in
capture events or by a fragmentation process during the collapse of a
single rotating protostar. In fact, 
Hartigan et al. (1994) have shown that about one third of wide 
pre-main-sequence binary pairs in their sample (projected separations
of 400--6000 AU) are not coeval, with the less massive star usually being
younger than the more massive star, suggesting that they are 
formed by capture or exchange (more likely) events.
On the other hand, close systems must almost 
certainly be primordial, although their exact origin is not yet well
understood. Several mechanisms have been suggested (fragmentation
during the late collapse, gravitational instabilities or even orbital
decay) for explaining the formation of these close primordial binaries.
Moreover, observations indicate that binary stars are seen since the earliest
stages of star formation, which suggests primordial mechanisms for their
origin (Harjunp\"a\"a et al. 1991; Reipurth \& Zinnecker 1993; Mathieu 1992,
1994, 1996).
In any case, observational results hint 
that star formation produces mostly binaries.
Further information about binaries in clusters can be found in Trimble (1980),
Abt (1983), Reipurth (1988), Mathieu (1989), Zinnecker (1989) and 
an extensive recent review in Bodenheimer et al. (1993). 
\hfill\par
Binaries in 
star clusters are
of chief importance both in observational and theoretical astrophysics. 
In some open clusters,
a certain number of stars appear above the cluster main-sequence turn-off
(blue stragglers); it
is currently interpreted (Wheeler 1979) as 
a result of stellar coalescence or extended 
main sequence life-times caused by mixing within the stars (Abt 1985).
Another popular explanation for the origin of blue stragglers 
may be due to binary mass transfer (McCrea 1964).
Also, runaway OB stars are interpreted as binaries escaping from star clusters
(De Cuyper 1982; Sutantyo 1982; Hills 1983; Gies \& Bolton 1986).
In addition, binaries
in open clusters are used to determine the distance scale (standard candles).
Binaries are also the key to explaining X-ray emissions from globular and old
open star clusters.
From a theoretical point of view, it is  
clear that primordial binaries (hereafter PBs) can dominate the early 
stages of the 
dynamical evolution of star clusters (see Heggie \& Aarseth 1992). 
\hfill\par
Although binaries are heavier than the mean mass in a cluster, 
their masses do not remain constant
throughout their life. Mass loss from stellar evolution, 
stellar collisions and exchange of mass 
from one component to another can alter 
significantly the binary orbit or even disrupt it. 
Primordial binaries 
which are too close will be modified or
even disrupted by mechanisms such as mass loss or 
mass transfer during the evolution 
of the primary from a main sequence star to a giant.
Because of their 
massive nature, binaries tend to occupy the inner regions of star clusters so 
a change in their physical parameters (mass is the most important) 
may affect significantly the entire dynamical evolution of the system. 
Their impact on the cluster evolution depends for the most part
on the coupling between the characteristic time-scale for mass
segregation in the cluster and that 
for stellar mass loss. If mass segregation 
can be reached before significant mass loss has occurred, these 
primordial binaries act as energy sources more or less continually 
until they become unstable by stellar evolution or escape. In the case of 
important stellar mass loss before equipartition, this situation 
may not be attained and the new evolved binaries may have no tendency to reach
the inner regions of the cluster because their masses have decreased in a
significant way. However, from a theoretical point of view, 
a smaller effect is expected than in the case of models without PBs but stellar
evolution, because the single stars are also losing mass, 
so binaries are in any case the most
massive objects.
\hfil\par
The role of stellar evolution for the dynamics of star clusters has been
of interest since the end of the seventies (Angeletti \& Giannone 1979, 1980;
Angeletti et al. 1980)
and there are a number of recent papers (Stod\'o{\l}kiewicz 1982, 1985;
Terlevich 1983, 1985, 1987; Applegate 1986; Weinberg \& Chernoff 1988;
Quinlan \& Shapiro 1990; de la Fuente Marcos 1993, 1996a (hereafter Paper II); 
Aarseth 1996b, c).
\hfil\par
The effect of PBs on the
dynamical evolution of clusters was studied before stellar 
evolution. The first paper on the subject was by Aarseth (1975) and 
it was followed by a number of papers (Aarseth 1980; Spitzer \& Mathieu 1980;
Giannone \& Molteni 1985;
McMillan et al. 1990, 1991a, b; Murphy et al. 1990; Gao et al. 1991; Hut et al. 1992;
Heggie \& Aarseth 1992; McMillan 1993; McMillan
\& Hut 1994; de la Fuente Marcos 1996b (hereafter Paper III)). In addition, 
Hills (1975) gave a semi-analytical discussion about this question, 
Goodman and Hut (1989) pointed out the importance of PBs for the evolution
of globular clusters and Leonard and Duncan
(1988, 1990) carried out $N$-body simulations with primordial binaries,
although their main emphasis was on escapers.
Summaries of most of these studies can be found in the introductory section
of McMillan et al. (1990), Gao et al. (1991) and Heggie and Aarseth (1992).
\hfil\par
On the contrary, the study of the interplay 
between stellar mass loss and PBs
on the evolution of star clusters has only very recently become 
of interest. Pols and Marinus
(1994) studied the binary stellar evolution in young open clusters 
using Monte--Carlo simulations, although their chief interest was in 
pure stellar evolution, and not in the dynamical one. 
Direct $N$-body calculations have been performed 
principally by Aarseth,
but only a few details have been published (Aarseth 1996b, c). He has found
that because of stellar evolution the fraction of binaries increases in 
the central regions
of rich star clusters ($N$ up to 10$^{4}$). This increase in the central binary 
fraction is because massive single stars evolve to low-mass stars.
\hfil\par
This paper is mainly devoted to study the interplay between the mass spectrum,
the PB fraction and the mass loss due to stellar evolution on the 
dynamical evolution of open star clusters. Moreover, we want 
to compare our results for the surviving binary fraction with observational
data for binaries in open clusters. We are mainly concerned with the binary 
type for trying to answer the question about the preferential  
type of surviving binaries in open clusters. It is to be expected that 
binaries with 
both components being low-mass stars (late spectral types) 
will be preferential survivors in 
rich open star clusters because the time-scale for cluster disruption is 
larger than their characteristic time-scale for significant mass loss 
due to stellar evolution. 
However, for poorly populated open clusters, the disruption time-scale can
be significantly smaller than the stellar evolution time even for moderately 
massive stars, so it should be possible to find binaries with massive components.
Also, we are interested in comparing the present
results with those from previous papers in this series (de la Fuente Marcos
1995 (hereafter Paper I); Paper II; Paper III) concerning the role of 
the initial mass function (hereafter IMF) 
on the dynamical evolution of cluster models. 
\hfil\par
We have performed five runs each for a total number of stars $N$ = 100, 250,
500, 750 with five different IMFs using direct 
integration methods. As in previous
papers, we use the same version of the Aarseth's code NBODY5 (Aarseth 1985; 
Aarseth 1996a). This code has become a standard in the field of star clusters
simulations. Written in FORTRAN, it consists of a fourth-order 
predictor-corrector integration scheme with individual time steps. It 
utilizes an Ahmad-Cohen (1973) neighbour scheme to facilitate calculation of 
the gravitational forces, and handles close encounters via two-, three-, 
four-, and chain regularization techniques (Kustaanheimo \& Stiefel 1965; 
Aarseth \& Zare 1974; Mikkola 1985; Mikkola \& Aarseth 1993). 
\hfil\par
All the calculations have been performed on a VAX 9000/210, running under
OpenVMS operating system, at the 
{\it Centro de Proceso de Datos (UCM, Moncloa, Madrid)}. This machine 
has one CPU and its peak performance is about 100 Mflops.  

\section{Cluster models}
In the present section all the physical features of the models 
we have performed are discussed.
This description follows closely that in the previous 
papers of this
series, although in this case all the astrophysical processes 
discussed above are included 
simultaneously.
\subsection{Initial Mass Function}
\begin{figure}[htbp]
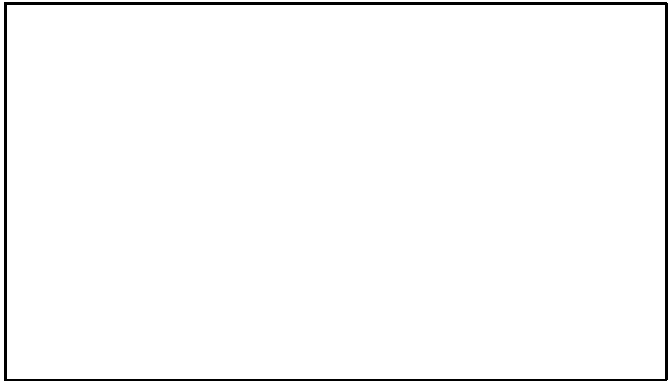

   \picplace{5cm}
   \caption{Distribution function for the modified Scalo IMF with the 
            binary correlation described in Paper III. Note the logarithmic 
            scale for both axes.}
      \label{IMFsca}
\end{figure}

The frequency distribution of stellar masses at birth is a fundamental 
parameter for studying the evolution of star clusters.
As in previous papers (Paper I; Paper II;
Paper III) several IMFs have been used 
in the calculations in order to generate an initial distribution of masses. 
We refer to Paper I for a full discussion
of the IMFs used. Models for Kroupa (Kroupa et al. 1993) and 
Scalo (Scalo 1986; Eggleton 1994) IMFs are used with the binary 
correlation described in Paper III (Eggleton 1995). 
In Fig. \ref{IMFsca} we see the modified Scalo IMF with the binary 
correlation used.
%
%                                                      One column figure
%-----------------------------------------------------------------IMF sca
%
%------------------------------------------------------------------------
%
For models with 
Salpeter (1955), Taff (1974) and Miller \& Scalo  
(1979) IMFs the two components of the binary have the same mass.
The five IMFs used in our calculations are summarized in Table \ref{IMFDATA}.       
%
%__________________________________________________________Table I (IMFs)
%
\begin{table}
   \caption{IMFs used in the calculations}
      \label{IMFused}
	 \begin{tabular}{lr}
	    \hline
	    Salpeter IMF   &  $\alpha$ = 2.35 \\
	    Taff IMF        &  $\alpha$ = 2.5 ($N \leq 100$) \\ 
			    &  $\alpha$ = 2.65 ($N > 100$) \\
	    Miller \& Scalo IMF      &  (Eggleton et al. 1989)$^{*}$ \\
	    Kroupa IMF     &  (Kroupa et al. 1993) \\
	    Scalo IMF   &  (Eggleton 1994)$^{+}$ \\
 	    \hline  
	 \end{tabular}
\begin{list}{}{}
\item[$^{\rm *}$] This IMF is a fit to the Miller \& Scalo's (1979) results.
\item[$^{\rm +}$] This IMF is a fit to the Scalo's (1986) results. 
\end{list} 

\label{IMFDATA}

\end{table}
%
%____________________________________________________________________________
%
%
\subsection{Stellar evolution}
As in Paper II we use the fitting functions by Eggleton et al. (1989) in order 
to obtain the stellar diminution of mass as a function of time for Population I
stars. 
These interpolation formulae are explained  
in the original Eggleton et al's paper and partially in Paper II. 
All the stars start on a zero-age main sequence (hereafter ZAMS) with a 
uniform composition of hydrogen, X = 0.7, helium, Y = 0.28, and metallicity, 
Z = 0.02. For computational convenience, the mass loss is implemented at
discrete intervals (accumulated mass diminution of 1 \%). Because of very 
different relative velocities, 
the actual mass loss is assumed to be instantaneous.
The expelled gas leaves the cluster without any effect on the cluster members.
We do not consider mass transfer processes in binaries, so the evolutionary 
times given by Eggleton's fitting formulae are not changed. 
Also, there is no presence of accretion disks around stars, so disk accretion 
can not affect the evolutionary tracks of the cluster stars.
There is only 
one way of changing the evolutionary time-scales for a given star; it can
be achieved if two stars collide forming a new object, blue-straggler or 
Thorne-\.Zytkow object (Thorne \& \.Zytkow 1977). 

\subsection{Binary fraction and other binary parameters}
The number of primordial binaries in the cluster is conveniently parameterized
by the binary fraction $f$ given by:
\begin{equation}
f \ = \ \frac{N_{b}}{N_{b} \ + \ N_{s}} \,,
\end{equation}
where $N_{b}$, $N_{s}$ are, respectively, the number of stars 
which are binaries and singles. As in Paper III, a binary fraction of $f$ = 1/3
is used in the calculations; so the overall multiplicity, defined as the 
ratio of the number of multiples to the total number of systems, is 0.33 
in our present models. Lower bounds from observational surveys in Star 
Formation Regions are 0.37 (Ophiuchus) and 0.55 (Taurus) (Simon et al. 1995). 
For a sample of stars in the solar neighbourhood, Duquennoy and Mayor (1991)
have found a fraction of 0.57, after correcting for observational bias. 
For comparing directly with models from previous
papers, the number of singles and binaries is chosen in such a way that the 
total number of stars (not objects) is 100, 250, 500 and 750 respectively.
This arbitrary choice has minor effects on the cluster dynamical evolution
as described in Paper III.
\hfil\par
In order to include a realistic initial binary population there are some other 
parameters to be determined in addition 
to the binary fraction. Our initial population of binaries is hard because 
they are of main dynamical importance for the evolution of star clusters.
A binary is defined to be hard or soft (Heggie 1980) depending on whether 
its binding energy is greater than or less than the local mean kinetic energy 
per star. As in Paper III, the
semi-major axis of the binaries is taken from a uniform distribution:
\begin{equation}
   a_{b}\:=\:a^{0}_{b}\:10^{-q} \,,
\end{equation}
where $a^{0}_{b}$ is an input parameter whose value is about $1/N$ in units
of $R_{\rm vir}$ (for a hard binary) and $q$ is equal to X log${\cal R}$. 
The virial radius, $R_{vir}$, is defined by
\begin{equation} 
R_{vir} \ = \ -G {\cal M}^{2} / 4 E \,, 
\end{equation}
where $E$ is the total energy of the system, excluding the binding energies
of any initial binaries, $G$ is the gravitational constant and ${\cal M}$
is the total mass of the cluster; ${\cal R}$ is an input parameter, and X is 
a random number uniformly distributed  in the interval $[0, 1]$. The spread in 
the energy of the binaries is given by the spread in semi-major axis and 
finally it is given by ${\cal R}$. Small values of ${\cal R}$ produce wide 
binaries, large values give close binaries. The value of ${\cal R}$ used in 
our simulations is the same that we used in Paper III, ${\cal R} = 10$.
A typical value for the initial semi-major axis of the binaries in our 
simulations is about 500 AU (the smaller value is about 70 AU and the maximum is 
about 3300 AU). The values for the semi-major axis 
in Table \ref{MODDAT} are the
upper cut-off for the semi-major axis distribution. 
Eccentricities are chosen from a random (thermalized) distribution (Jeans 1929)
and the same is done for the pericentre, node and inclination. The mass ratio 
for the PBs in all the models is 1/2, excepting those in which the binary 
correlation has been used.

\subsection{Main features of the models}
All the models (for the same $N$) have the same sequence of random numbers 
for generating initial conditions; spherical symmetry and constant density are 
assumed, with the ratio of total kinetic and potential energy fixed at 0.25. 
Another characteristics common to all models are random and isotropic initial 
velocities. Also, in all the cases the cluster 
suffers mass loss due to escape of stars.
A star escapes and is removed from the calculation when its distance 
from the cluster centre is greater than twice the tidal radius. The tidal 
radius is given by the classical expression $r_{t} \approx (\frac
{G {\cal M}}{T_{1}})^{1/3}$ where $T_{1}$ is defined as a function
of $A$ and $B$  
Oort constants of galactic rotation: 
$T_{1} \ = \ 2 \omega [ (- A - B) - \omega]$, where $\omega$ is the rotational 
velocity of the star cluster around the galactic nucleus.
As in previous papers, the galactic gravitational field is 
introduced in the models as described in Terlevich (1987).
Moreover, we ignore the effect of field stars on the dynamical 
evolution of the cluster due to their high relative velocities (see 
discussion in Paper II). 
\hfil\par					  
We use a standard and consistent set of units throughout, except where 
explicitly noted. All lengths are measured in parsecs.  
Times are measured in terms 
of the initial half-mass crossing time, defined by 
\begin{equation}
   T_{cr} = G {\cal M}^{5/2} / ( -2 E)^{3/2} \,.  
\end{equation}
It represents the time taken for a typical star, moving with 
velocity $< v^{2} >^{1/2}$, to cross the virial diameter (2 $R_{vir}$). 
As in previous papers of this series we adopt a 
consistent tidal field for the models which have the same mean 
stellar density. A typical value for the 
mean mass density in a real cluster is 1.3 
${\cal M}_{\odot} pc^{-3}$ (Lohmann 1971, 1976a, 1976b, 1977a, 1977b) 
with a range of 0.5-3.2 ${\cal M}_{\odot} pc^{-3}$ 
for his sample of open clusters. 
These values are typical for evolved clusters,
so that a larger mean density of $\simeq 12 
{\cal M}_{\odot} pc^{-3}$ is adopted for the whole cluster (Table \ref{MODDAT})
as considered young and not evolved.
 All the models do not have the same maximum 
 (${\cal M}_{\rm max}$) and minimum (${\cal M}_{\rm min}$)
 masses for generating the IMF. The Salpeter,Taff and Scalo IMFs have
 ${\cal M}_{\rm max}$ = 15.0 ${\cal M}_{\odot}$ and ${\cal M}_{\rm min}$ = 0.1 
${\cal M}_{\odot}$ but the Kroupa and Eggleton IMFs use an algorithm that 
changes upper and lower limits for masses (see Table \ref{MODDAT}). 
Hence models for the Salpeter and Taff IMFs have the value of mean 
density quoted above but the others have $\simeq 6 \ {\cal M}_{\odot} 
pc^{-3}$ (if using the same mean stellar mass, this will give the initial 
virial radius for different N).
%
%____________________________________________________________Table 2 (CHAR)
%
\begin{table*}

\caption[ ]{Main characteristics of the models}

\begin{flushleft}

\begin{tabular}{llllllllllll}
\hline\noalign{\smallskip}
$MODEL$ & $IMF^{*}$ & $N^{\bullet}$ & $N_{b}$ & $a_{b}^{0 \ \times}$ 
        & ${\cal M}_{\rm max}^{\dagger}$ 
	& ${\cal M}_{\rm min}^{\dagger}$ & $<{\cal M}>^{\dagger}$
        & $R_{0}^{\ddagger}$  
	& $r_{t \ 0}^{\star}$ & $<R>_{0}^{\diamond}$ & $T_{d}^{\odot}$ \\
\noalign{\smallskip}
\hline\noalign{\smallskip}
I & SA & 100 & 25 & 0.0160 & 15.0 & 0.1 & 1.0 & 1.26 & 5.96 & 0.78 & 30.4 (74) \\
II & TA & 100 & 25 & 0.0160 & 15.0 & 0.1 & 1.0 & 1.26 & 5.95 & 0.78 & 47.4 (116) \\
III & MS & 100 & 25 & 0.0160 & 15.0 & 0.1 & 0.7 & 1.26 & 5.39 & 1.06 & 91.9 (261) \\
IV & KR & 100 & 25 & 0.0160 & 5.5 / 1.8 & 0.2 / 0.1 & 1.2 / 0.4 & 1.26 & 5.16 & 1.05 & 111.7 (335) \\
V & SC & 100 & 25 & 0.0160 & 5.8 / 2.0 & 0.3 / 0.1 & 1.3 / 0.4 & 1.26 & 5.37 & 1.10 & 89.9 (256) \\
VI & SA & 250 & 62 & 0.0116 & 15.0 & 0.1 & 1.0 & 1.71 & 8.11 & 1.40 & 118.1 (288) \\
VII & TA & 250 & 62 & 0.0116 & 15.0 & 0.1 & 1.0 & 1.71 & 8.11 & 1.32 & 184.6 (450) \\
VIII & MS & 250 & 62 & 0.0116 & 15.0 & 0.1 & 0.6 & 1.71 & 7.03 & 1.37 & 171.8 (517) \\
IX & KR & 250 & 62 & 0.0116 & 5.5 / 4.8 & 0.2 / 0.1 & 0.9 / 0.4 & 1.71 & 6.75 & 1.66 & 177.5 (568) \\
X & SC & 250 & 62 & 0.0116 & 5.8 / 5.1 & 0.2 / 0.1 & 1.0 / 0.5 & 1.71 & 7.20 & 1.62 & 217.8 (634) \\
XI & SA & 500 & 125 & 0.0123 & 15.0 & 0.1 & 1.0 & 2.15 & 10.19 & 2.00 & 205.1 (498) \\
XII & TA & 500 & 125 & 0.0123 & 15.0 & 0.1 & 1.0 & 2.15 & 10.19 & 1.91 & 280.0 (680) \\
XIII & MS & 500 & 125 & 0.0123 & 15.0 & 0.1 & 0.6 & 2.15 & 8.75 & 1.98 & 204.4 (625) \\
XIV & KR & 500 & 125 & 0.0123 & 5.5 / 5.1 & 0.2 / 0.1 & 0.9 / 0.4 & 2.15 & 8.51 & 2.02 & 270.1 (859) \\
XV & SC & 500 & 125 & 0.0123 & 5.8 / 5.4 & 0.2 / 0.1 & 1.1 / 0.5 & 2.15 & 9.05 & 1.98 & 262.4 (763) \\
XVI & SA & 750 & 187 & 0.0081 & 15.0 & 0.1 & 1.0 & 2.47 & 11.68 & 2.30 & 329.5 (804) \\
XVII & TA & 750 & 187 & 0.0081 & 15.0 & 0.1 & 1.0 & 2.47 & 11.68 & 2.30 & 387.5 (945)  \\
XVIII & MS & 750 & 187 & 0.0081 & 15.0 & 0.1 & 0.6 & 2.47 & 9.98 & 2.32 & 162.6 (503) \\
XIX & KR & 750 & 187 & 0.0081 & 5.5 / 5.2 & 0.2 / 0.1 & 0.9 / 0.4 & 2.47 & 9.73 & 2.27 & 227.0 (729) \\
XX & SC & 750 & 187 & 0.0081 & 5.8 / 12.9 & 0.2 / 0.1 & 1.0 / 0.6 & 2.47 & 10.30 & 2.27 & 369.8 (1061) \\
\noalign{\smallskip}
\hline

\end{tabular}

\end{flushleft}
%
%------------------------------------------------------------------------
%
\begin{list}{}{}
\item[$^{*}$] SA Salpeter IMF, TA Taff IMF, 
                  MS Miller \& Scalo IMF, KR Kroupa IMF, SC Scalo IMF.
\item[$^{\bullet}$] Total number of stars ($N_{s} \ + \ 2 \ N_{b}$).
\item[$^{\times}$]  Semi-major axis for PBs in pc.
\item[$^{\dagger}$] In ${\cal M_{\odot}}$. For KR and SC models 
                    (Binary / Single).
\item[$^{\ddagger}$] Initial virial radius in pc. 
\item[$^{\star}$] Initial tidal radius in pc.
\item[$^{\diamond}$] Initial half-mass radius in pc.
\item[$^{\odot}$] Disruption time in scaled units (in Mys).
\end{list}

\label{MODDAT}

\end{table*}
%____________________________________________________________________________
%

Table \ref{MODDAT} gives the disruption time for our present models.
If we compare these values with those from Paper III we observe that for 
$N$ = 100 the disruption time is now smaller only for models with 
Salpeter, Taff and Miller \& Scalo IMFs. However, for models with 
Kroupa or Scalo IMFs the disruption time is increased even comparing with
Paper II. The main difference between the two groups of IMFs is the maximum
mass. For the first group, the upper cut-off for the mass spectrum is 15 
solar masses but this value is significantly smaller for the second group, 
so the reason
for this behaviour is clearly due to the massive stars. 
It seems that the supernova events
in the core destabilize the cluster in a very efficient way for poorly 
populated clusters. Two supernova 
events are enough to provide the energy source which disrupts the cluster.
The acceleration of disruption as compared with Paper II is very significant
for models with power-law IMF, however for Miller \& Scalo IMF the disruption time
in the current models is greater than the respective ones for Paper II (due to 
an upper cut-off in the mass spectrum). 
For models with $N$ = 250, the same trend pointed out above is observed for
power-law models with regard to Paper III. 
\hfil\par
On the contrary, realistic 
IMFs show greater disruption times than those of Paper II and III. For 
$N$ = 500, the behaviour of the disruption time is 
very similar to $N$ = 250. For $N$ = 750, 
power-law models have disruption times greater than those of Paper III but
a bit smaller than those of Paper II. For Miller \& Scalo and Kroupa IMFs
the disruption times are smaller than those of Paper III and Paper II, 
however for the Scalo IMF the disruption time is greater than those from 
Paper II and III. This suggests a complex interplay between the IMF, 
mass loss from stellar evolution and primordial binaries. For poor clusters
the massive stars in PBs dominate the cluster evolution. For
rich clusters, the interplay between power-law IMFs, PBs and stellar
evolution seems to increase the 
cluster life-time in relation to models with PBs but no stellar evolution.
For intermediate population ($N$ = 250, 500), power-law models with
all realistic features
disrupt earlier than their respective models 
without mass loss. However, there is no
clear interpretation for this behaviour. The increased life-time in some 
models for $N$ = 750 can be explained by the formation of temporarily stable 
multiple systems (in some cases, hierarchical triple systems) in the cluster
remnant; for $N$ = 500, a certain number of these systems are also formed in
some models. These systems delay the disruption of the 
evolved cluster because some of them are long-lived. Formation 
of such systems depends strongly on the fraction of PBs, the cluster 
membership and the parameter ${\cal R}$. 
Larger $N$ and PB fraction promote an increased probability of
formation of such multiple systems.

\section{Characteristic quantities} 
In this section we compare all of the different runs,
concentrating, as in previous papers, on two representative 
diagnostic parameters. The first one measures
the degree of dynamical evolution of the entire cluster. This quantity is
 called the evolution modulus and is defined by (von Hoerner 1976) 
\begin{equation}
  W = log ( R_{\rm h} / R_{\rm c} ) \,,  
\end{equation}
where $R_{\rm h}$ is the half-mass radius and $R_{\rm c}$ is the core radius. 
The core radius defined by Casertano and Hut (1985) is
\begin{equation}
  R^{(CH)}_{\rm core} = \Sigma_{i} \ \rho_{i} \ R_{i} / 
			\Sigma_{i} \ \rho_{i} \,. 
\end{equation}  
Here, $R_{i}$ is the distance from star $i$ to the density center, defined 
as the density-weighted centroid of the system, the density $\rho_{i}$ is 
determined by the distance $R_{6, i}$ to star $i$'s sixth nearest neighbour:
\begin{equation}
  \rho_{i} = M_{6, i} / R^{3}_{6, i} \,,
\end{equation}
where $M_{6, i}$ is the total mass lying within distance $R_{6, i}$ of star 
$i$ (excluding the mass of star $i$ itself), and the sum is taken over all 
stars in the system.
The second parameter is the escape rate $dN / dt$ which describes the 
disruption rate of the system. 

\subsection{Evolution Modulus} 
This quantity has 
the advantage that it can be obtained directly from observable properties.
%
%                                                      One column figure
%-----------------------------------------------------------------E_mod VI
\begin{figure}[htbp]
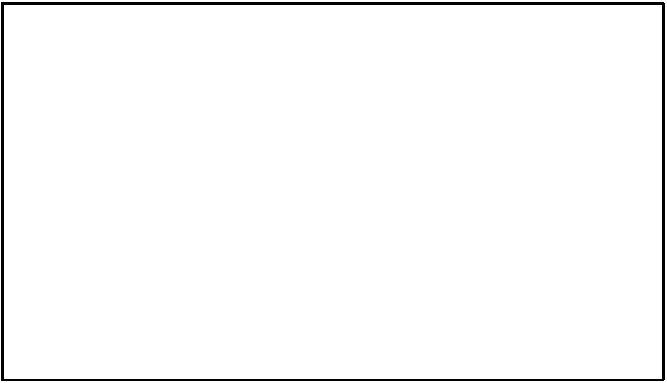

   \picplace{5cm}
   \caption{Evolution modulus for Salpeter's models with $N$ = 100, 
	    250, 500, 750. In all figures, time is given in units
            of the half-mass initial crossing time in the current 
            system.}
      \label{FigEvoModVI}
\end{figure}
%
%------------------------------------------------------------------------
%
%
%                                                      One column figure
%-----------------------------------------------------------------E_mod VII
\begin{figure}[htbp]
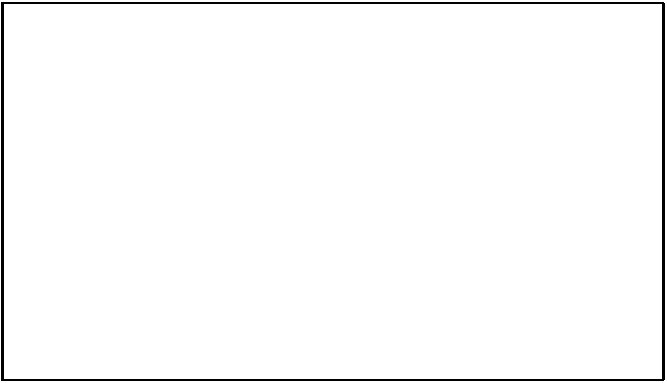

   \picplace{5cm}
   \caption{Evolution modulus for Taff's models with $N$ = 100, 250,  
	    500, 750.}
      \label{FigEvoModVII}
\end{figure}
%
%------------------------------------------------------------------------
%
%
%                                                      One column figure
%-----------------------------------------------------------------E_mod VIII
\begin{figure}[htbp]
   \picplace{5cm}
   \caption{Evolution modulus for Miller \& Scalo's 
            models with $N$ = 100, 250, 
	    500,750.}
      \label{FigEvoModVIII}
\end{figure}
%
%------------------------------------------------------------------------
%
%
%                                                      One column figure
%-----------------------------------------------------------------E_mod IX
\begin{figure}[htbp]
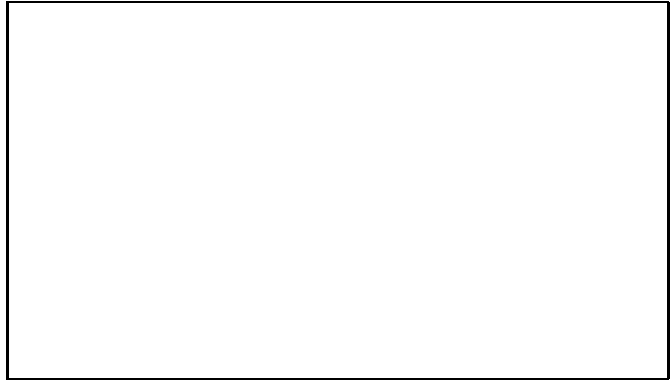

   \picplace{5cm}
   \caption{Evolution modulus for Kroupa's models with $N$ = 100, 250,  
	    500, 750.}
      \label{FigEvoModIX}
\end{figure}
%
%------------------------------------------------------------------------
%
%
%                                                      One column figure
%-----------------------------------------------------------------E_mod X
\begin{figure}[htbp]
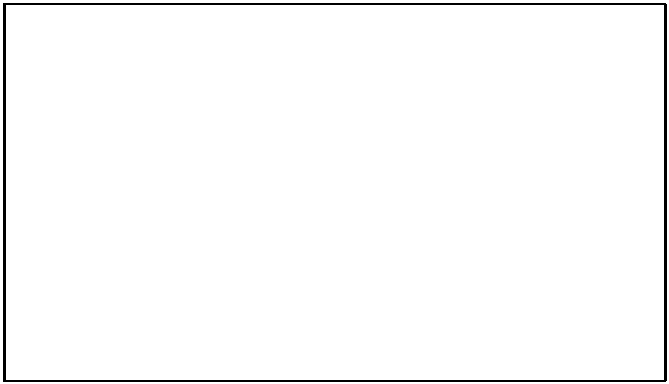

   \picplace{5cm}
   \caption{Evolution modulus for Scalo's models with $N$ = 100,
	    250, 500, 750.}
      \label{FigEvoModX}
\end{figure}
%
%------------------------------------------------------------------------
%
Figs. \ref{FigEvoModVI}-\ref{FigEvoModX} show the behaviour 
of $W$ as a function of time for all the models. 
If we compare the present figures with those from Paper II we observe
significant differences. 
\hfil\par
For Salpeter IMF models, the behaviour of $W$ is very
similar for all $N$ in Paper II, however the inclusion of PBs shows great
differences, depending on $N$, for our present models. 
First, the evolution time-scales (in scaled
units) are very different depending on $N$, accelerating the dynamical 
evolution in poor clusters. For $N$ = 100, the cluster is disrupted even 
having $W > 0$. For $N$ = 250 the evolution is also 
accelerated but for $N$ = 500 the evolutive time-scales are almost the same 
as without PBs. The evolution is even slower for $N$ = 750. 
As before, this trend can be explained by the formation of temporarily bound 
multiple systems (triple and quadruple).
The comparison 
between our present results for $W$ and those from Paper III also presents  
several differences. The evolution of $W$ is slower for greater $N$ but
for smaller $N$ it is nearly similar. The mean value of $W$ is now significantly
smaller because the halo is less extended than in the case of models
with PBs but no stellar evolution.
\hfil\par
For Taff IMF models, the behaviour of $W$ is roughly similar to that 
presented in Paper II except in the case of $N$ = 100. In this case, 
the cluster evolution is accelerated in a very significant way, but 
the cluster is completely disrupted after reaching $W$ = 0. The behaviour 
is different if comparing with Paper III because the core-halo structure 
seems to be stabilized by stellar mass loss as compared with models
with PBs but no mass loss.
\hfil\par
Models with Miller \& Scalo IMF show practically similar 
behaviour to those presented in Paper II. The only main differences 
appear for early stages of the cluster evolution where the mean value
of $W$ is greater for models without PBs. This suggests 
that the haloes formed in models with PBs and stellar evolution are
less extended. As regards comparison with models without stellar 
evolution we also observe features suggesting that the 
interplay between mass loss and PBs produces less energetic haloes than
in the case of models with PBs but no stellar evolution.
\hfil\par
For models with Kroupa IMF, we observe that the evolution is very
different from that presented in Paper II. For large $N$ the evolution
is slowed down but for $N$ = 100 the evolution is accelerated although less
than for power-law models. The evolution of $W$ is affected even at 
early stages of the cluster evolution, when the stellar mass loss is 
not yet dominant. If comparing with Paper III, the evolution of the
cluster seems to be slightly accelerated.
\hfil\par
Finally, for Scalo IMF we observe that with regard to Paper II the 
evolution for models with $N \geq 250$ is significantly slowed down, but
speeding up a bit for smaller $N$. As regards Paper III, the 
ratio core-halo (in size) seems to be more stable for models with mass loss and 
the overall evolution is slowed down.
\hfil\par
We observe that the inclusion of stellar evolution in models with PBs 
affects their evolution considerably but in a very uncertain way. However,
it is clear that the changes are very IMF dependent. The interpretation
of the results for our present models in relation to models with PBs but 
no stellar evolution is very unclear in comparison with the conclusions 
we obtained with regard to models without PBs but with or without mass loss
due to stellar evolution. 

\subsection{The escape rate} 
\begin{figure}[htbp]
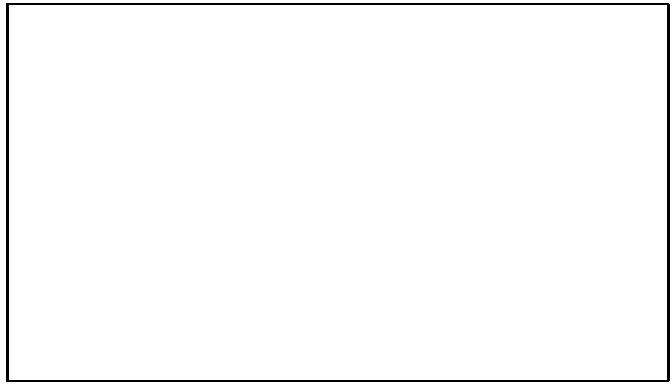

   \picplace{5cm}
   \caption{Evolution of the cluster population with time for models with  
	    $N$ = 100.}
      \label{Population_100}
\end{figure}

\begin{figure}[htbp]
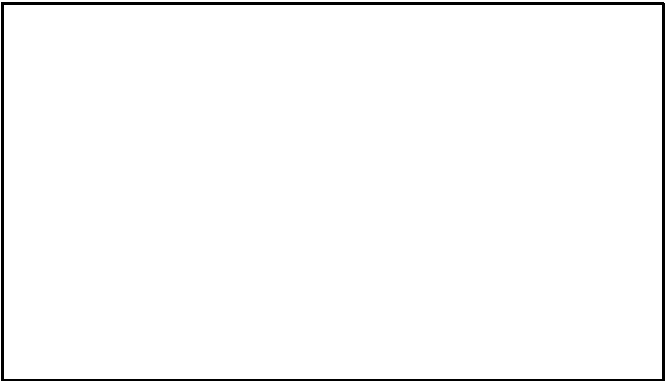

   \picplace{5cm}
   \caption{Evolution of the cluster population with time for models with 
	    $N$ = 750.}
      \label{Population_750}
\end{figure}

This is another important quantity for the 
cluster evolution. Figures \ref{Population_100}-\ref{Population_750} 
show the number
of cluster stars as a function of time. 
%
%                                                      One column figure
%----------------------------------------------------------Population N = 100
%
%------------------------------------------------------------------------
%
%
%                                                      One column figure
%----------------------------------------------------------Population N = 250
\begin{figure}[htbp]
   \picplace{5cm}
   \caption{Evolution of the cluster population with time for models with  
	    $N$ = 250.}
      \label{Population_250}
\end{figure}
%
%------------------------------------------------------------------------
%
%
%                                                      One column figure
%----------------------------------------------------------Population N = 500
\begin{figure}[htbp]
   \picplace{5cm}
   \caption{Evolution of the cluster population with time for models with 
	    $N$ = 500.}
      \label{Population_500}
\end{figure}
%
%------------------------------------------------------------------------
%
%
%                                                      One column figure
%----------------------------------------------------------Population N = 750
%
%------------------------------------------------------------------------
%
For $N$ = 100 the inclusion of stellar evolution in models with PBs 
distinguishes clearly between models with power-law IMFs and realistic ones.
In comparison with figures for the escape rate without stellar evolution,
the disruption of power-law models is accelerated but in the case of 
realistic IMFs it is slowed down. Even for Scalo IMF we observe an increase
in the duration of the stage in which close encounters are dominant. The 
almost exponential behaviour that we observe in some models in Paper III
disappears because most of the escapers are not due to distant encounters.
As regards comparison with Paper II the evolution of the escape rate is 
very different so we suggest that the escape mechanism induced by 
binaries with stellar evolution is different from the dominant one in 
clusters with mass loss due to stellar evolution but no PBs.
\hfil\par
The same trend is also observed for $N$ = 250 but it is not so clear as in 
the case of $N$ = 100. Models with Taff and Miller \& Scalo IMFs show 
a nearly similar behaviour as in Paper III. 
\hfil\par 
For $N$ = 500 the behaviour of the escape rate is very different from that
presented in Paper III. Models with massive stars (power-law IMFs) lose
their distinctive features and accelerate their disruption rates 
(in scaled units). For the other IMFs the figures are nearly similar to
their respective ones without PBs. With regard to comparison with Paper III, the 
escape rate is reduced by including stellar evolution.
\hfil\par
For $N$ = 750 the inclusion of stellar evolution in models with PBs produces
a significant change in the escape rate for power-law IMFs.
Their escape rate show a slow-down with respect to those from Paper III.
However, models with realistic IMFs, except for Scalo IMF, look very 
similar to those from Paper III. With regard to comparison with Paper II results,
the escape rate is slowed down for models with Miller \& Scalo, Kroupa 
and power-law IMFs but is accelerated for Scalo IMF.
\hfil\par
As we can see, the interpretation of the results is difficult in most of the cases.
Only for $N$ = 100 can we do this in an easy way. Clearly the differences 
induced in these models are affected by the massive stars. The stellar 
evolution in models with small $N$ and PBs is the dominant mechanism for the 
dynamical evolution of open clusters. For most of the models with small $N$ 
we found in Paper III almost exponential decay in the escape rate  
which suggests an evaporative (by distant encounters) dominant escape mechanism, 
however now we find a change in the shape of the curve. The reason seems 
to be due to the halo extension. Models including stellar evolution seem to show
less extended haloes than models without, so the number of stars available
for leaving the system in a smooth way is smaller. This trend almost 
dissapears for models with $N$ = 250 but it can still be observed preferentially at 
early stages of the cluster evolution. For larger $N$ the explanation 
is not very clear, also we note less extended haloes but the results
depend strongly on the IMF. In any case the dominant mechanism for 
the escape rate at early stages of the star cluster evolution seems to be by 
close encounters.   

\section{Evolution of PB population}
\begin{figure}[htbp]
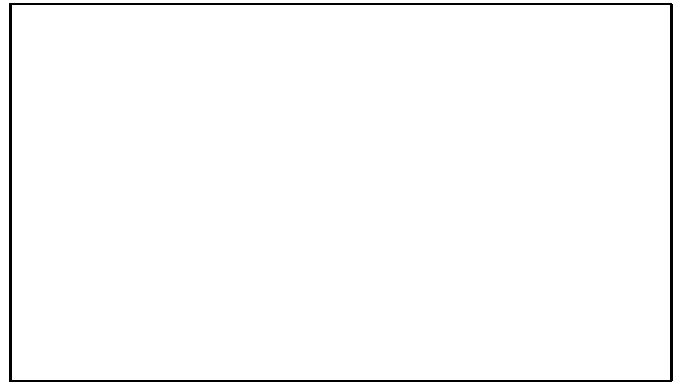

   \picplace{5cm}
   \caption{Evolution of the PB population as a function of time 
            for models with 
	    $N$ = 100.}
      \label{B1}
\end{figure}

\begin{figure}[htbp]
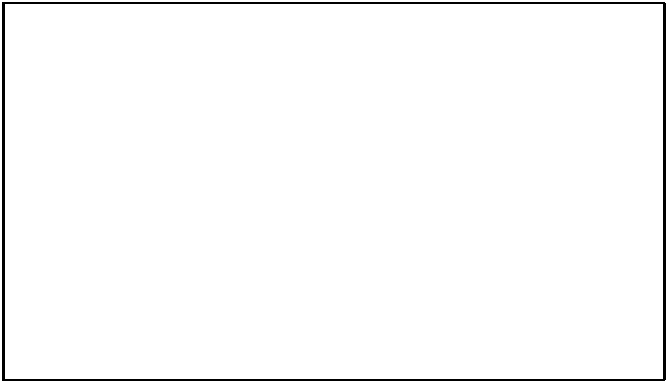

   \picplace{5cm}
   \caption{Evolution of the PB population as a function of time 
            for models with 
	    $N$ = 750.}
      \label{B4}
\end{figure}

This section is devoted to study the evolution of the 
PBs in our present models. We are mainly interested in comparing with results
from Paper III. In Figs. \ref{B1}-\ref{B4} we see the percentage of
primordial binaries in the cluster as a function of time.
%
%                                                      One column figure
%----------------------------------------------------------PBs   N = 100
%
%------------------------------------------------------------------------
%
%
%
%                                                      One column figure
%----------------------------------------------------------PBs   N = 250 
\begin{figure}[htbp]
   \picplace{5cm}
   \caption{Evolution of the PB population as a function of time 
            for models with 
	    $N$ = 250.}
      \label{B2}
\end{figure}
%
%------------------------------------------------------------------------
%
%
%
%                                                      One column figure
%----------------------------------------------------------PBs   N = 500
\begin{figure}[htbp]
   \picplace{5cm}
   \caption{Evolution of the PB population as a function of time 
            for models with 
	    $N$ = 500.}
      \label{B3}
\end{figure}
%
%------------------------------------------------------------------------
%
%
%
%                                                      One column figure
%----------------------------------------------------------PBs   N = 750
%
%------------------------------------------------------------------------
%
%

From the figures, the first thing to note is the 
life time of the primordial binary population. For poorly populated
models ($N$ = 100), we obtain that the life 
time of the PBs are significantly shorter 
than for Paper III models. Moreover, the evolution of the surviving 
fraction of the PB population for models
with power-law IMFs seems to be clearly different from that observed
for models with realistic IMFs. The diminution in the number of PBs 
is more rapid because the evolution of the cluster itself is quicker.
The reason is due to the supernova events. In Salpeter and Taff models
two supernova events have occurred; as a result the cluster 
disintegration and the escape or disruption of PBs are accelerated.
For $N$ = 250, some differences are observed as regard models from
Paper III. Salpeter models show a significant acceleration in the 
diminution rate of the PB fraction caused, as before, by the supernova
explosions (2 for these models). 
However, the Taff model shows a nearly similar behaviour as
in Paper III; the same can be observed for realistic IMFs except for 
the Scalo model. For $N$ = 500 the almost exponential decrease in the 
percentage of surviving binaries which appears in models from Paper III 
disappears for the majority of the present models. Initially the same
linear diminution is observed, but later the diminution rate slows down
by a very significant value. For some models in Paper III the percentage 
of surviving PBs at the cluster mean life time is smaller than 30 \% but
now this value reaches almost 50 \% for power-law IMF models. 
Models with $N$ = 750 also show differences. In Paper III, all the models
(excluding the monocomponent one) have a nearly similar behaviour but now
models with Salpeter, Taff and Scalo IMFs show almost the same behaviour
but the others do not. For the former models the diminution of PBs is almost
linear and smaller than for Miller \& Scalo and Kroupa IMFs. From Table 
\ref{MODDAT}, we see that models for Salpeter, Taff and Scalo IMFs 
have more massive stars so this may be the reason for this behaviour.
In average, our models show that PBs are retained preferentially; this result
has also been found by Aarseth (1996d).
\hfil\par
Our present considerations suggest that the membership is 
a main factor for retaining PBs; models with increasing $N$ show 
smaller diminution rates for the percentage of surviving PBs. Also, the 
IMF has a main role in the evolution of PBs; for poorly populated clusters
massive stars in binaries control the evolution of the entire cluster. 
For small open clusters the presence of massive stars in binaries can 
accelerate their disruption but for more populated star clusters the effect
is to the contrary. 
%
%__________________________________________________Table 3 (Binary fraction)
%
\begin{table}

\caption[ ]{Binary fraction in the core and in the whole cluster}

\begin{flushleft}

\begin{tabular}{lllllll}
\hline\noalign{\smallskip}
$MODEL$ & $f^{0}_{c}$ & $f^{0}_{t}$ & $f^{h}_{c}$ & $f^{h}_{t}$ &
          $f^{e}_{c}$ & $f^{e}_{t}$ \\ 
\noalign{\smallskip}
\hline\noalign{\smallskip}
I & 0.31 & 0.33 & 0.00 & 0.35 & 0.00 & 0.13 \\
II & 0.33 & 0.33 & 0.00 & 0.28 & 0.00 & 0.57 \\
III & 0.63 & 0.33 & 0.80 & 0.28 & 0.36 & 0.57 \\
IV & 0.45 & 0.33 & 0.33 & 0.27 & 0.18 & 0.57 \\
V & 0.47 & 0.33 & 0.27 & 0.23 & 0.09 & 0.22 \\
VI & 0.33 & 0.33 & 0.32 & 0.31 & 0.25 & 0.50 \\
VII & 0.35 & 0.33 & 0.00 & 0.31 & 0.00 & 0.83 \\
VIII & 0.30 & 0.33 & 0.33 & 0.31 & 0.27 & 0.57 \\
IX & 0.33 & 0.33 & 0.17 & 0.29 & 0.09 & 0.57  \\
X & 0.36 & 0.33 & 0.36 & 0.39 & 0.33 & 0.71 \\
XI & 0.36 & 0.33 & 0.75 & 0.34 & 0.19 & 0.57 \\
XII & 0.37 & 0.33 & 0.42 & 0.31 & 0.09 & 0.38 \\
XIII & 0.34 & 0.33 & 0.50 & 0.32 & 0.09 & 0.83 \\
XIV & 0.36 & 0.33 & 0.33 & 0.28 & 0.08 & 0.63 \\
XV & 0.27 & 0.33 & 0.33 & 0.28 & 0.25 & 0.33 \\
XVI & 0.34 & 0.33 & 0.00 & 0.35 & 0.00 & 0.63 \\
XVII & 0.33 & 0.33 & 0.00 & 0.28 & 0.00 & 0.80 \\
XVIII & 0.32 & 0.33 & 0.47 & 0.32 & 0.33 & 0.50  \\
XIX & 0.37 & 0.33 & 0.55 & 0.25 & 0.13 & 0.22  \\
XX & 0.36 & 0.33 & 0.55 & 0.30 & 0.29 & 0.56  \\
\noalign{\smallskip}
\hline

\end{tabular}

\end{flushleft}
%
%------------------------------------------------------------------------
%
\label{BINF}

\end{table}
%____________________________________________________________________________
%

\begin{figure}[htbp]
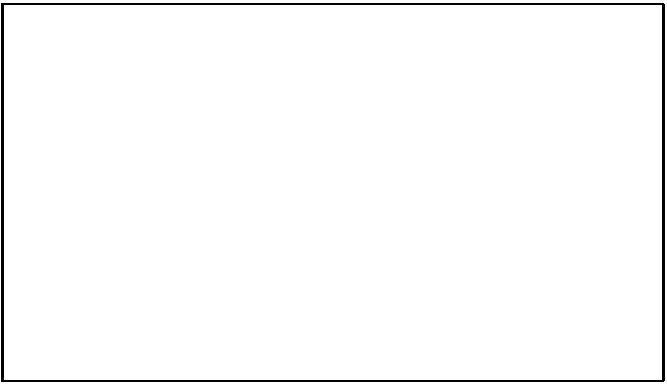

   \picplace{5cm}
   \caption{Evolution of the total binary fraction as a function of time 
            in years for models with 
	    $N$ = 750. In all the models an initial decay is observed
            but it increases in most of the models when the cluster 
            population has reached values smaller than one third 
            of the initial.}
      \label{bifra}
\end{figure}

An interesting parameter to study is the binary fraction, $f$, both in the
core and the whole cluster. Table \ref{BINF} gives the core and total 
binary fractions for all the models at three selected
times in the cluster life: at time equal to zero ($f^{0}_{c}$, $f^{0}_{t}$),
at the time in which the stellar population is a half of the initial one
($f^{h}_{c}$, $f^{h}_{t}$) and at the end of the simulation 
($f^{e}_{c}$, $f^{e}_{t}$). In this Table we only consider the 
primordial binaries because they are the only hard ones; 
in our present models binaries which are formed dynamically
are all soft and short lived. Some quick conclusions arise
from the present values of the binary fraction. First, for power-law models
there is a strong tendency for binaries to underpopulate the core on a short
time scale. Except for  $N$ = 500 all the power-law models show a smaller
binary fraction in the core at the half cluster life. The explanation is due to
the supernova events located in the core. They dominate the evolution,  
driving the
binaries outside the core (values for 
the total binary fraction are nearly similar
for all the models at the cluster half life). 
A typical behaviour of the total binary fraction can be observed in Fig.
\ref{bifra}. In all the models an initial decay is observed but the total
binary fraction increases in most of the cases when the cluster population has
decreased significantly. It suggests a preferential escape
of the single stars in clusters with a fraction of PBs.
However, the evolution of the
core binary fraction with time is extremely irregular. Its behaviour is 
extremely chaotic although almost in all the models an initial increase is
observed as in Aarseth (1996b, c). This effect is greater for rich models 
and realistic IMFs. Second, the value of the total binary fraction towards the
end of our simulations for $N \leq 250$ is 0.57 in many of the models; it 
is the value currently accepted for the binary fraction in the solar 
neighbourhood (Duquennoy \& Mayor 1991). This value appears preferentially for
models with Miller \& Scalo and Kroupa IMFs so it is possible to suggest that
there probably exists a link between the size of the cluster in which stars 
are born, their IMF and the binary fraction in a certain region of space.
%
%                                                      One column figure
%----------------------------------------------------------fPBs   N = 750
%
%------------------------------------------------------------------------
%
%

\section{Velocity distribution of the escaping stars}
In this section we analyze the velocity distribution of the escaping stars,
both singles and binaries. The main emphasis is on comparing present results
with previous ones for models without PBs and mass loss, with mass loss but
no PBs and with PBs but no mass loss. Most of the results that appear in 
this section were not discussed previously in the papers of this series.
\hfil\par
Stars are considered as escapers and they are removed from the calculations
when they reach a distance from the cluster centre greater  
than twice the tidal radius. At that moment, we note 
the velocity of the star (if single) or of the centre of mass 
(if binary). The escape boundary depends on the Galactic parameters so 
the obtained velocities are higher that those obtained from the 
theoretical expression ($2\:<v>$); many of our velocities have excess escape
energy.
The values of the mean escape velocity, the standard deviation in 
the mean, the minimum escape velocity and the maximum escape velocity can 
be seen in Table \ref{ESCV}.
%
%____________________________________________________________Table 3 (ESCV)
%
\begin{table}

\caption[ ]{Escape velocity for the models$^{\star}$}

\begin{flushleft}

\begin{tabular}{lllll}
\hline\noalign{\smallskip}
$MODEL$ & $<v_{esc}>$ & $\sigma_{m}$ & $v_{esc}^{min}$ & $v_{esc}^{max}$ \\ 
\noalign{\smallskip}
\hline\noalign{\smallskip}
I & 0.935 & 0.070 & 0.261 & 4.450 \\
II & 0.866 & 0.062 & 0.413 & 4.426 \\
III & 0.822 & 0.056 & 0.309 & 3.511 \\
IV & 0.699 & 0.044 & 0.000 & 2.413 \\
V & 0.710 & 0.052 & 0.223 & 3.137  \\
VI & 1.190 & 0.065 & 0.251 & 7.551 \\
VII & 1.270 & 0.051 & 0.434 & 5.992 \\
VIII & 0.996 & 0.044 & 0.000 & 5.768 \\
IX & 0.858 & 0.030 & 0.270 & 3.994  \\
X & 1.049 & 0.049 & 0.210 & 6.915 \\
XII & 1.419 & 0.050 & 0.000 & 11.261 \\
XIII & 1.456 & 0.057 & 0.000 & 18.427 \\
XIV & 1.140 & 0.031 & 0.245 & 5.050 \\
XV & 1.051 & 0.023 & 0.241 & 4.726 \\
XVI & 1.211 & 0.036 & 0.264 & 7.819 \\
XVII & 1.558 & 0.030 & 0.345 & 6.960 \\
XVIII & 1.544 & 0.036 & 0.345 & 8.687 \\
XIX & 1.141 & 0.024 & 0.267 & 6.386  \\
XX & 1.158 & 0.026 & 0.000 & 7.988  \\
XXI & 1.246 & 0.030 & 0.000 & 11.080  \\
\noalign{\smallskip}
\hline

\end{tabular}

\end{flushleft}
%
%------------------------------------------------------------------------
%
\begin{list}{}{}
\item[$^{\star}$] All the data in km/s.
\end{list}

\label{ESCV}

\end{table}
%____________________________________________________________________________
%
%
We observe two clear tendencies: first, the mean escape velocity increases
with $N$ and second, the dispersion of velocities decreases when $N$ 
increases. This is to some extent connected with the physical scaling in the
models.
Moreover, the greater mean escape velocities are always for 
models with Salpeter and Taff IMFs, so it depends on the IMF. This is to be
expected because the masses of these models are greater than for the 
other IMFs. As we can see, when $N$ increases the probability of 
high velocity escapers grows; we observe that in all cases the maximum 
escape velocity is considerably higher than the mean escape velocity. 
The origin of these high velocity escapers is connected with binary-binary
interactions so the high escape velocities must also be connected with 
binaries of a certain size. 
\hfil\par
As regards comparison with results from previous models, we observe 
significant differences. Considering results from models of Paper III, the
mean velocity of the escapers is reduced in most of the present models.
This tendency is clear for clusters with simple power-law IMFs. However, 
realistic IMFs show a more complex behaviour. For $N$ = 100, the same tendency
as for old IMFs is observed (except for Kroupa IMF). For $N$ = 200, the 
mean velocity is larger for the current models. For $N$ = 500, this is only
observed for Miller \& Scalo IMF. For $N$ = 750, the same tendency is observed
for all the models. Concerning the theoretical implications of these experimental
results, it seems that stellar evolution of massive stars is the main
process which explains the observed differences. Mass loss in binaries
promotes the disruption of a certain number of such systems for models 
with simple power-law IMFs so the global results suggest the number of 
initial binaries would have been smaller. On the other hand, present values for the mean 
velocity of the escapers are larger than those of Paper II, so the presence 
of PBs increase the velocity of the escapers (due to close encounters 
between binaries and singles) for models with mass loss.
\hfil\par
As regards the relation of velocities of escaping stars with their masses,
we observe a bimodal distribution. Most of the escaping particles have 
velocities not much larger than the mean but a few per cent has velocities 
greater than three times the mean. This suggests two 
processes for the escape of cluster stars. Low velocity escapers are 
generated by evaporation, i.e. gradual increase of the velocity because of 
distant encounters and high velocity escapers are produced by close 
encounters between singles or, more frequently in models with PBs, 
between singles and binaries or binary-binary encounters.

\section{Mass loss in binaries}
In order to obtain the change of mass as a function of time, NBODY5 
includes the fast fitting functions by Eggleton et al. (1989) for
Population I stars. For convenience, mass loss is implemented at 
discrete intervals when the accumulated contribution reaches 1 percent.
The actual mass loss is assumed to be instantaneous, with no further
effect on the cluster members or binary companions. Approximate
energy conservation can be achieved by performing appropriate corrections
to the total potential energy. In the version of NBODY5 used, the 
binding energy is corrected by the term $\Delta \ {\cal M} / r$ where
$\Delta \ {\cal M}$ is the mass loss and $r$ is the 
binary separation. For close binaries the mass loss is always
at apocentre, hence the effect will be smallest. When the mass loss is
large, the binary is usually disrupted.
\hfil\par
The loss of mass from a binary system is a very complicated dynamical 
problem. Stellar mass loss causes changes in the orbital elements 
of the binary. The problem has been mainly studied by Hadjidemetriou
(1963, 1966, 1968).
\hfil\par
The orbital period can increase or decrease secularly 
depending upon the mass-flow conditions. The simplest case is where 
mass is lost isotropically from the system. This situation is assumed
in our stellar mass loss events.
In our models, the vast majority of massive binaries are disrupted 
during mass loss events, so the preferential binary survivors are 
low mass binaries in which the mass loss rate is sufficiently slow to 
permit a quasi-stable evolution.
\hfil\par
By Kepler's third law,
\begin{equation}
\frac{4 \ \pi^{2} \ a^{3}}{P^{2}} \ = \ G \ {\cal M}
\end{equation}
where $P$ is the binary period. 
It gives, for a constant semi-major axis $a$, the following relation between
the change in period $\Delta \ P$ for a mass loss $\Delta \ {\cal M}$:
\begin{equation}
\frac{\Delta P}{P} \ = \ - \frac{\Delta {\cal M}}{2 \ {\cal M}} \,. \label{MM}
\end{equation}
From Eq. (\ref{MM}) an abrupt change in the period can be achieved by one
component losing mass in an eruptive outburst, with the lost 
material ejected at a high speed. From our simulations, it is found that
the evolution of surviving low mass PBs (not exchanged) 
with the two components of the same mass
can be described almost exactly by Eq. (\ref{MM}) with small 
deviations due to the perturbers; i.e. the semi-major axis remains almost
constant during most of the time when the mass loss rate is not too high. 
The evolution of massive binaries 
is very complicated because of mass loss. The majority of these binaries 
are destroyed during the first few crossing times. Their products, mainly
white dwarfs, appear sometimes as members of another binary. These binaries
containing collapsed objects have a mass ratio of nearly one.   
\hfil\par
The above formulae can be used for estimating the changes in the period 
for an unperturbed binary. The phenomena is more complex if we consider
that the binary has several (in some case many) perturbers. In this case
mass loss events near apocentre can easily disrupt the binary because
the perturbing force is of the same order as the force between binary
components. This depends mainly on the binary semi-major axis; binaries 
with small semi-major axis are like single stars from a dynamical point
of view, so the mass loss events are less destructive for them. On
the other hand stellar mass loss in clusters with PBs can favour the survival
of certain kinds of binaries; binaries with both low-mass primary and 
secondary do not suffer a significant mass loss before their possible
escape from the cluster. There is observational evidence
(Verbunt et al. 1994) that favour the above hypothesis; most X-ray
sources in the old open cluster M67 are RS CVn binaries.  
These are active binary systems in which the primary is a star with a 
spectral type F or G in the main sequence or subgiant (luminosity class 
V or IV) and the secondary is a bit cooler and usually more massive and evolved 
with spectral type K and class IV. Their orbital periods are usually in the 
range 1--14 days. In models for $N$ = 750 we have found several binaries 
of this spectral type in the remnant, typically with a period of a few 
hundred years although in other models (not detailed here) with 
$N$ = 1500 and 500 primordial binaries the typical value for the period
is about 8 days. In any case, the mean value for the periods of the 
surviving binaries depends on the initial period distribution. 
 
\section{Evolution of the stellar content}
As pointed out above all our models start with the stars on the ZAMS.
This assumption implies some questionable hypotheses. First, it assumes that
the stellar formation in the cluster took place by a single burst. On the
other hand, it means there is no preferential stellar mass for the beginning
of the star formation. The second of these premises is arguable
in the light of some observational discoveries (Herbig 1962; Iben \& Talbot
1966; Cohen \& Kuhi 1979; Adams et al. 1983; Strom 1985). They suggest that
low-mass stars form first and over a long time scale. However, Strom (1985)
cautioned that these conclusions rested heavily on the theoretical 
pre-main-sequence tracks. As pointed out in the excellent
review by Zinnecker et al. (1993), low-mass stars acquire their final
mass first, but will then take a long time to reach the ZAMS; high-mass
stars take longer to accumulate their mass but they evolve rapidly
onto the ZAMS. Indeed, the data available are consistent with simultaneous
formation of stars of all masses (Stahler 1985; Schroeder \& Comins 1988).
In spite of these questionable initial hypotheses, the evolution of 
the stellar content of our models is analyzed in the present section.
\hfil\par
We want to study specially the 
destiny of the collapsed objects in our models. As we see from Table
\ref{MODDAT}, some of our models (preferentially with power-law IMFs) have 
stars with enough mass to finish their nuclear life in a supernova event.
From our current knowledge of stellar evolution, after such an episode the 
final product may be a neutron star (called pulsar if its radio emission
can be detected from the Earth) or a black hole. The last kind of collapsed 
object has not been introduced in the stellar evolution routines used in
our present models. As regards the formation of neutron stars, it is 
considered in our models. There are a significant number of neutron stars 
(pulsars) detected in globular clusters, but at present there is no evidence 
for these collapsed objects in open clusters. In our models there are no 
surviving neutron stars. They escape from the cluster shortly after their 
formation. It is because after the formation of a pulsar, it suffers a 
kick velocity due to a strong non-symmetric mass loss. Although models with 
greater $N$ can retain a population of neutron stars, our present models 
are not able to keep the neutron stars. The time-scale for leaving 
the cluster for a newly-born neutron star in our models is a few kyr, so
it can be considered as an explanation for the null population of neutron
stars observed in open clusters. However, it is to be expected from models 
with greater $N$ ($N$ = 1500 in models not presented here) that there is a higher 
(but low) chance of detecting a neutron star in a highly populated open
cluster (maybe M67 is a good candidate to achieve this).
\hfil\par
On the contrary, white dwarfs remain in the cluster for many crossing times in 
most of the models. We have even found a few models in which there are 
one or two white dwarfs among the members of the cluster remnant. 
However, there are some differences between the life times for 
white dwarfs in our models depending both on the IMF and cluster membership.

\section{Open Cluster Remnants}
In this section we consider the composition of the final cluster remnant
in our models. In Paper III, we noted that the final remnants 
had a common feature, a high binary richness. This is 
also observed in the present models but the fraction of purely primordial 
binaries remaining in the final object resulting from the cluster disruption
has decreased. Also we observe that the binaries with two components of 
the same mass are the preferential survivors. In the majority of the models
the number of remaining binaries is three or four (we stop the simulations 
when the cluster population is $N$ = 10). From the results of models for 
Papers I and II, we also find a certain number of binaries in the cluster
remnant (usually 1 or 2). These binaries are formed dynamically, not 
primordials so one of them is hard and the others (if any) are soft. In our
present models all of the surviving binaries are hard.
\hfil\par
The final binary population shows distinctive characteristics depending
on the cluster richness. For models with $N \leq 250$ the surviving binaries
do not have a preferential ratio between the masses of the binary components.
We find binaries with both components being low-mass stars (${\cal M} \leq 
0.8 M_{\odot}$) or both massive or one of them massive and the other not.
However, for rich models the surviving binaries are always with almost equal-mass
components. In a few cases one component is a giant and the other is a
low-mass star with a spectral type typically in the range M0-M5. When the two
components are on the ZAMS both the primary and secondary have spectral type
in the range K5-M5. In general, the stellar content of the open cluster
remnants is mostly ZAMS stars and in a few models there appears a white dwarf or
a red giant, especially for rich clusters. In most of the models the binary 
fraction in the remnant is significantly larger than the initial one;
Aarseth (1996d) has also found this effect. In a model with 10,500 stars and
500 binaries, the remnant contains 10 binaries and 39 stars.
\hfil\par
Although binaries in cluster remnants are primordial, i.e. at least one
of their components was in a PB  at the beginning of the simulation, not all
the surviving binaries are purely primordial. For rich clusters we find
a certain fraction of exchanged binaries (1 or 2 per model). For small 
clusters it is rare to find an exchanged binary because the cluster disrupts
faster so the probability of an exchange during the cluster life is reduced.
\hfil\par
The explanation of the above results rests on the time-scales for stellar
evolution. For clusters with small $N$, the cluster disrupts before massive
stars have started to leave the ZAMS, so we can find binaries with massive
primaries in the cluster remnant. However for rich clusters, the disruption 
time-scale is greater than the characteristic time-scale 
for significant mass loss
from massive stars, so binaries with massive primaries could be disrupted 
or ejected during mass loss events. 
The observational test of our results is very difficult in the case 
of rich clusters remnants. The detection of a small population 
(a few tens) of faint stars
most of them less luminous than our Sun is a big challenge for the biggest
optical telescopes. Moreover, the detection of the binaries in these 
rich cluster remnants is even more difficult. However, there is some light from
this dark picture because stars with such spectral types are expected to be
strong radio and/or X-rays sources which can be detected with synthesis 
techniques or by instruments on satellites. 
The search of open cluster remnants 
has been considered by Lod\'en (1977, 1979, 1980), who 
analyzed very loose or star-poor clusterings in a large survey of the
Southern Milky Way. Thousands of these objects were found and classified
into 4 sets. One of them is formed by extremely small and star-poor 
clusters which he suggested as possibly cluster remnants.
The frequency of these objects in his sample is about 20 \% but he considers
that it can be significantly greater.

\section{Conclusions}

The main conclusions from this work can be summarized as follows:

1.- The inclusion of stellar evolution in cluster models with a fraction
    of PBs affects the overall dynamical evolution of the cluster 
    in a uncertain way which depends strongly on the cluster richness and
    the IMF. 

2.- The stellar evolution in clusters with small population and power-law IMF 
    accelerates their disruption in a very significant way.

3.- The escape velocity increases with the cluster richness but the dispersion
    decreases when $N$ increases. Sometimes, a star can leave the cluster
    with high velocity.

4.- The final cluster remnant is very rich in binaries, frequently purely
    primordial but its composition depends strongly on the initial 
    cluster population because of the interplay between the time-scales 
    for cluster disruption and stellar mass loss. Binaries in remnants of 
    poor clusters do not have any special feature in their components but 
    binaries in rich cluster remnants have usually almost the same mass for 
    their components and have late spectral types. Collapsed objects are
    almost always absent from open cluster remnants.

\begin{acknowledgements}

I am very grateful to my Ph. D. thesis supervisor Dr. Sverre J. Aarseth
for his generous help, useful comments, for providing his computer code
and for sharing some results before publication. 
The author is also grateful to Dr. P. P. Eggleton for the use of his binary
correlation before publication and his IMF fits.
I thank the Department of Astrophysics of Universidad Complutense de Madrid
for providing excellent computing facilities. 
All the computations and most of the data analysis 
were made on the computers of the 
Universidad Complutense of Madrid, and I thank the computing staff 
(specially J. Palero and E. Lezcano) for their 
help during this stage. 
The author gratefully acknowledges the hospitality of the Institute of
Astronomy (Cambridge, U.K.) during several visits when work on the papers
of this series was carried out.
This research has made use of SIMBAD (operated
at CDS, Strasbourg, France), BDA (operated at the Institute of Astronomy, 
University of Lausanne, Switzerland) and ADS (operated by NASA) databases.

\end{acknowledgements}


\begin{thebibliography} {}

   \bibitem{} Aarseth S. J., 1975,
      in: Dynamics of Stellar Systems, 
      ed.\ A. Hayli, D. Reidel Publ., p.\ 57
     
   \bibitem{} Aarseth S. J., 1980,
      in: Star Clusters,
      ed.\ J. E. Hesser, D. Reidel Publ., p.\ 325      

   \bibitem{} Aarseth S. J., 1985,
      in: Multiple Time Scales,
      eds.\ J. U. Brackbill, B. I. Cohen,
      Academic Press, New York, p.\ 377

   \bibitem{} Aarseth S. J., 1996a,
      in: Computational Astrophysics: Gas Dynamics and Particle Methods,
      eds.\ W. Benz, J. Barnes, E. M\"uller, N. Norman,
      Springer Verlag, in press  

   \bibitem{} Aarseth S. J., 1996b,
      in: The Origins, Evolution and Destinies of Binary Stars in Clusters,
      eds.\ E. F. Milone, J.-C. Mermilliod, P. Hut, 
      ASP Conference Series v.\ 90, p.\ 423

   \bibitem{} Aarseth S. J., 1996c,
      in: Dynamical Evolution of Star Clusters--Confrontation of Theory and
      Observations,
      eds.\ P. Hut, J. Makino, 
      Kluwer, Dordrecht, p.\ 161

   \bibitem{} Aarseth S. J., 1996d,
      private communication

   \bibitem{} Aarseth S. J., Zare K., 1974,
      Celest. Mech. 10, 185

   \bibitem{} Abt H. A., 1983,
      ARA\&A 21, 343

   \bibitem{} Abt H. A., 1985,
      ApJL 294, L103

   \bibitem{} Adams M. T., Strom S. E., Strom K. M., 1983,
      ApJ 53, 893

   \bibitem{} Ahmad A., Cohen L., 1973,
      J. Comput. Phys. 12, 389

   \bibitem{} Angeletti L., Giannone P., 1979,
      A\&A 74, 57

   \bibitem{} Angeletti L., Giannone P., 1980,
      A\&A 85, 117

   \bibitem{} Angeletti L., Dolcetta R., Giannone P., 1980,
      Ap\&SS 69, 45

   \bibitem{} Applegate J. H., 1986,
      ApJ 301, 132

   \bibitem{} Bodenheimer P., Ruzmaikina T., Mathieu R. D., 1993,
      in: Protostars and Planets III, 
      eds.\ E. H. Levy, J. I. Lunine, The University of Arizona Press, p.\ 367

   \bibitem{} Brandner W., Alcal\'a J. M., Kunkel M., Moneti A., Zinnecker H., 1996,
      A\&A 307, 121

   \bibitem{} Casertano S., Hut P., 1985,
      ApJ 298, 80

   \bibitem{} Cohen M., Kuhi L. V., 1979,
      ApJS 41, 743

   \bibitem{} De Cuyper J.-P., 1982,
      in: Binary and Multiple Stars as Tracers of Stellar Evolution,
      ed.\ Z. Kopal and J. Rahe,
      D. Reidel Publ., p.\ 417

   \bibitem{} Duquennoy A., Mayor M., 1991,
      A\&A 248, 485

   \bibitem{} Eggleton P. P., 1994,
      private communication

   \bibitem{} Eggleton P. P., 1995,
      in preparation 

   \bibitem{} Eggleton P. P., Fitchett M. J., Tout C. A., 1989,
      ApJ 347, 998
   
   \bibitem{} de la Fuente Marcos R., 1993,
      Unpublished Dissertation, Univ.\ Complutense de Madrid

   \bibitem{} de la Fuente Marcos R., 1995,
      A\&A 301, 407 (Paper I)

   \bibitem{} de la Fuente Marcos R., 1996a,
      A\&A 308, 141 (Paper II)

   \bibitem{} de la Fuente Marcos R., 1996b,
      A\&A 314, 453  (Paper III)

   \bibitem{} Gao B., Goodman J., Cohn H., Murphy B., 1991,
      ApJ 370, 567

   \bibitem{} Ghez A. M., Neugebauer G., Matthews K., 1993,
      AJ 106, 2066

   \bibitem{} Ghez A. M., Emerson J. P., Graham J. R., Meixner M., Skinner C., 1994,
      ApJ 434, 707

   \bibitem{} Giannone G., Molteni , 1985,
      A\&A 142, 321

   \bibitem{} Gies D. R., Bolton C. T., 1986,
      ApJSS 61, 419

   \bibitem{} Goodman J., Hut P., 1989,
      Nature 339, 40

   \bibitem{} Hadjidemetriou J. D., 1963,
      Icarus 2, 440

   \bibitem{} Hadjidemetriou J. D., 1966,
      Icarus 5, 34

   \bibitem{} Hadjidemetriou J. D., 1968,
      Ap\&SS 1, 336

   \bibitem{} Harjunp\"a\"a P., Liljestr\"om T., Mattila K., 1991,
      A\&A 249, 493

   \bibitem{} Hartigan P., Strom K. M., Strom S. E., 1994,
      ApJ 427, 961

   \bibitem{} Heggie D. C., 1980,
      in: Globular Clusters,
      eds.\ D. Hanes, B. Madore, 
      Cambridge University Press, p.\ 281

   \bibitem{} Heggie D. C., Aarseth S. J., 1992,
      MNRAS 257, 513

   \bibitem{} Herbig G. H., 1962,
      ApJ 135, 736   

   \bibitem{} Hills J. G., 1975,
      AJ 80, 1075

   \bibitem{} Hills J. G., 1983,
      ApJ 267, 322

   \bibitem{} von Hoerner S., 1976,
      A\&A 46, 293

   \bibitem{} Hut P., McMillan S., Romani R. W., 1992,
      ApJ 389, 527

   \bibitem{} Iben I., Talbot R. J., 1966,
      ApJ 144, 968
 
   \bibitem{} Jeans J. H., 1929,
      Astronomy and Cosmogony, 2nd edn, Cambridge University Press;
      also Dover, New York, 1961

   \bibitem{} Kroupa P., Tout, C.A., Gilmore G., 1993,
      MNRAS 262, 545

   \bibitem{} Kustaanheimo P., Stiefel E. L., 1965,
      J. Reine Angew. Math. 218, 204

   \bibitem{} Leinert Ch., Haas M., Richichi A., Zinnecker H., Mundt R., 1991,
      A\&A 250, 407

   \bibitem{} Leinert Ch., Zinnecker H., Weitzel N., et al., 1993,
      A\&A 278, 129

   \bibitem{} Leonard P. J. T., Duncan M. J., 1988,
      AJ 96, 222

   \bibitem{} Leonard P. J. T., Duncan M. J., 1990,
      AJ 99, 608

   \bibitem{} Lod\'en L. O., 1977,
      A\&AS 29, 31

   \bibitem{} Lod\'en L. O., 1979,
      A\&AS 36, 83

   \bibitem{} Lod\'en L. O., 1980,
      in: Star Clusters,
      ed.\ J. E. Hesser, D. Reidel Publ., p.\ 121      

   \bibitem{} Lohmann W., 1971,
      Astron. Nachr. 292, 193

   \bibitem{} Lohmann W., 1976a,
      Ap\&SS 41, 27

   \bibitem{} Lohmann W., 1976b,
      Ap\&SS 45, 27

   \bibitem{} Lohmann W., 1977a,
      Ap\&SS 47, 447

   \bibitem{} Lohmann W., 1977b,
      Ap\&SS 51, 173

   \bibitem{} Mathieu R. D., 1989,
      in: Highlights of Astronomy,
      ed.\ J.-P. Swings, D. Reidel Publ., vol.\ 8, p.\ 111

   \bibitem{} Mathieu R. D., 1992,
      in: IAU Colloq. 135, Complementary Approaches to Double and Multiple
      Star Research, ed.\ H. A. McAlister, W. I. Hartkopf (Chelsea: ASP), p.\ 30

   \bibitem{} Mathieu R. D., 1994,
      ARA\&A 302, 465

   \bibitem{} Mathieu R. D., 1996,
      in: The Origins, Evolution and Destinies of Binary Stars in Clusters,
      eds.\ E. F. Milone, J.-C. Mermilliod,
      ASP Conference Series, v.\ 90, p.\ 231      

   \bibitem{} McCrea W. H., 1964,
      MNRAS, 128, 147

   \bibitem{} McMillan S., 1993,
      in: Dynamics of Globular Clusters,
      eds.\ S. Djorgovski, G. Meylan,
      ASP Conference Series, p.\ 171

   \bibitem{} McMillan S., Hut P., 1994,
      ApJ 427, 793

   \bibitem{} McMillan S., Hut P., Makino J., 1990,
      ApJ 362, 522
   
   \bibitem{} McMillan S., Hut P., Makino J., 1991a,
      ApJ 372, 111

   \bibitem{} McMillan S., Hut P., Makino J., 1991b,
      in: The Formation and Evolution of Star Clusters, 
      ed.\ K. Janes, ASP Conference Series, p.\ 421

   \bibitem{} Mikkola S., 1985,
      MNRAS 215, 171

   \bibitem{} Mikkola S., Aarseth S. J., 1993,
      Celest. Mech. Dyn. Astron., 57, 439

   \bibitem{} Miller G. E., Scalo J. M., 1979,
      ApJS 41, 513

   \bibitem{} Murphy B. W., Cohn H. N., Hut P., 1990,
      MNRAS 245, 335

   \bibitem{} Padgett D. L., Strom S. E., Edwards S., et al., 1996,
      in: Disks and Outflows around Young Stars, 
      eds.\ J. Staude, S. V. W. Beckwith, Springer-Verlag, in press
 
   \bibitem{} Pols O. R., Marinus M., 1994,
      A\&A 288, 475

   \bibitem{} Prosser C. F., Stauffer J. R., Hartmann L., et al., 1994,
      ApJ 421, 517

   \bibitem{} Quinlan G. D., Shapiro S. L., 1990,
      ApJ 356, 483

   \bibitem{} Reipurth B., 1988,
      in: Formation and Evolution of Low Mass Stars, 
      eds.\ A. K. Dupree, M. T. V. T. Lago, 
      D. Reidel Publ., p.\ 305

   \bibitem{} Reipurth B., Zinnecker H., 1993,
      A\&A 278, 81

   \bibitem{} Richichi A., Leinert Ch., Jameson R., Zinnecker H., 1994,
      A\&A 287, 145

   \bibitem{} Salpeter E. E., 1955,
      ApJ 121, 161

   \bibitem{} Scalo M. J., 1986,
      Fundam. Cosmic Phys. 11, 1

   \bibitem{} Schroeder M. C., Comins N. F., 1988,
      ApJ 326, 756

   \bibitem{} Simon M., Chen W. P., Howell R. R., Benson J. A., Slowik D., 1992, 
      ApJ 384, 212

   \bibitem{} Simon M., Ghez A. M., Leinert Ch., 1993, 
      ApJ 408, L33
   
   \bibitem{} Simon M., Ghez A. M., Leinert Ch. et al., 1995, 
      ApJ 443, 625

   \bibitem{} Spitzer L., Mathieu R. D., 1980, 
      ApJ 241, 618

   \bibitem{} Stahler S. W., 1985,
      ApJ 293, 207

   \bibitem{} Stod\'o{\l}kiewicz J. S., 1982,
      Acta Astr. 32, 63

   \bibitem{} Stod\'o{\l}kiewicz J. S., 1985,
      in: Dynamics of Star Clusters,
      eds.\ J. Goodman, P. Hut,
      D. Reidel Publ., p.\ 361  

   \bibitem{} Strom S. E., 1985,
      in: Protostars and Planets II, 
      eds.\ D. C. Black, M. S. Matthews, The University of Arizona Press, p.\ 17

   \bibitem{} Sutantyo W., 1982,
      in: Galactic X-Ray Sources,
      eds.\ P. W. Sanford, P. Laskarides, and J. Salton,
      Wiley, p.\ 27

   \bibitem{} Taff L. G., 1974,
      AJ 79, 11

   \bibitem{} Terlevich E., 1983,
      Ph. D. thesis, University of Cambridge

   \bibitem{} Terlevich E., 1985,
      in: Dynamics of Star Clusters,
      eds.\ J. Goodman, P. Hut,
      D. Reidel Publ., p.\ 471

   \bibitem{} Terlevich E., 1987,
      MNRAS 224, 193

   \bibitem{} Thorne K. S., \.Zytkow A. N., 1977,
      ApJ 212, 832

   \bibitem{} Trimble V. L., 1980,
      in: Star Clusters,
      ed.\ J. E. Hesser, 
      D. Reidel Publ., p.\ 259

   \bibitem{} Verbunt F., Johnston H., Hasinger G., Belloni T., Bunk W., 1994,
      in: Symposium on Interacting Binary Stars in conjunction with the 105
      Meeting of the Astronomical Society of the Pacific, eds.\ A. E. Shafter,
      p.\ 244

   \bibitem{} Weinberg M. D., Chernoff D. F., 1988,
      Bull. Am. Astron. Soc. 20, 964

   \bibitem{} Wheeler J. C., 1979,
      ApJ 234, 569 

   \bibitem{} Zinnecker H., 1989,
      in: ESO Workshop on Low Mass Star Formation 
      and Pre-Main Sequence Objects,
      ed.\ B. Reipurth, 
      European Southern Obs., p.\ 447

   \bibitem{} Zinnecker H., McCaughrean M. J., Wilking B. A., 1993,
      in: Protostars and Planets III, 
      eds.\ E. H. Levy, J. I. Lunine, The University of Arizona Press, p.\ 429

\end{thebibliography}
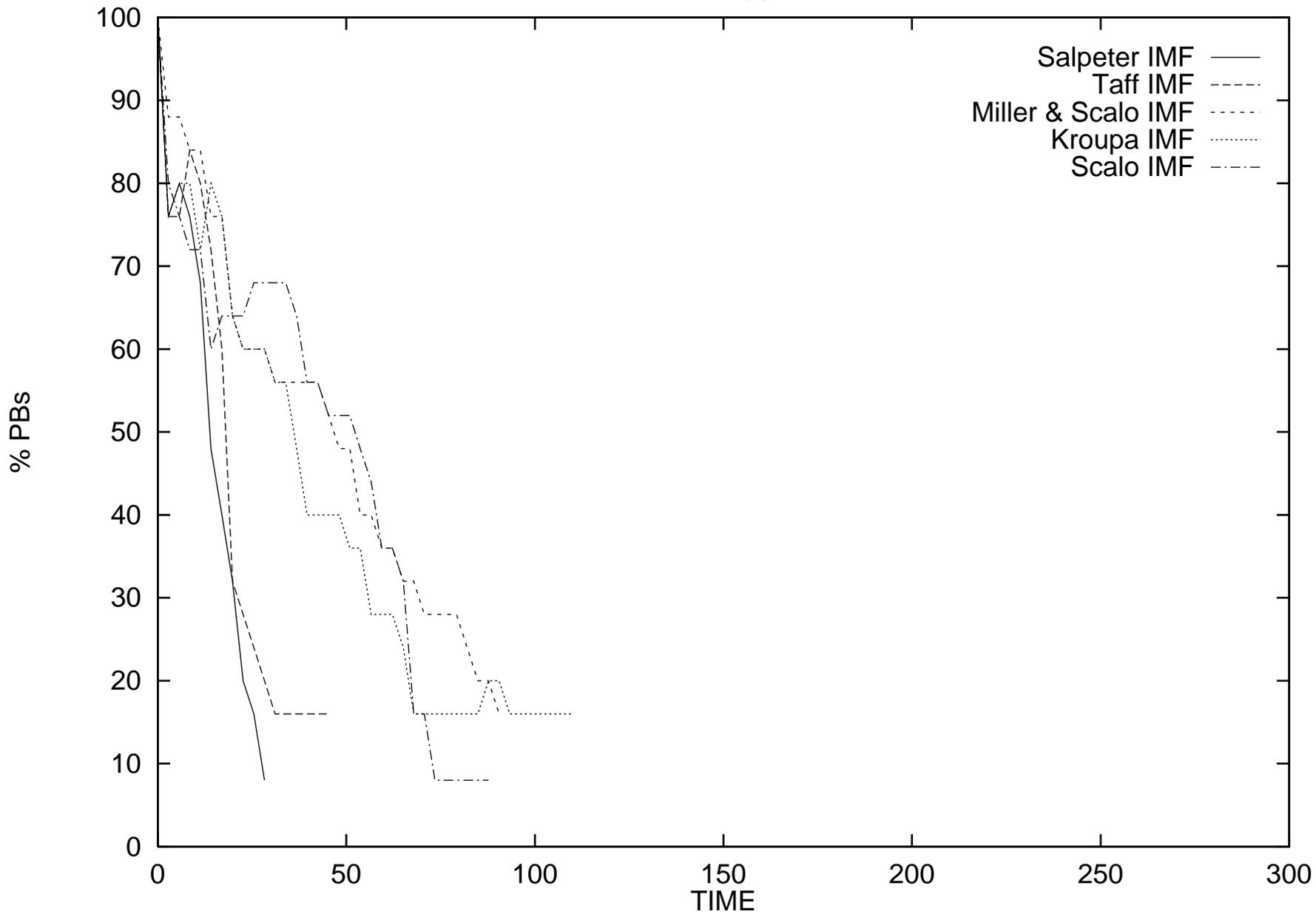
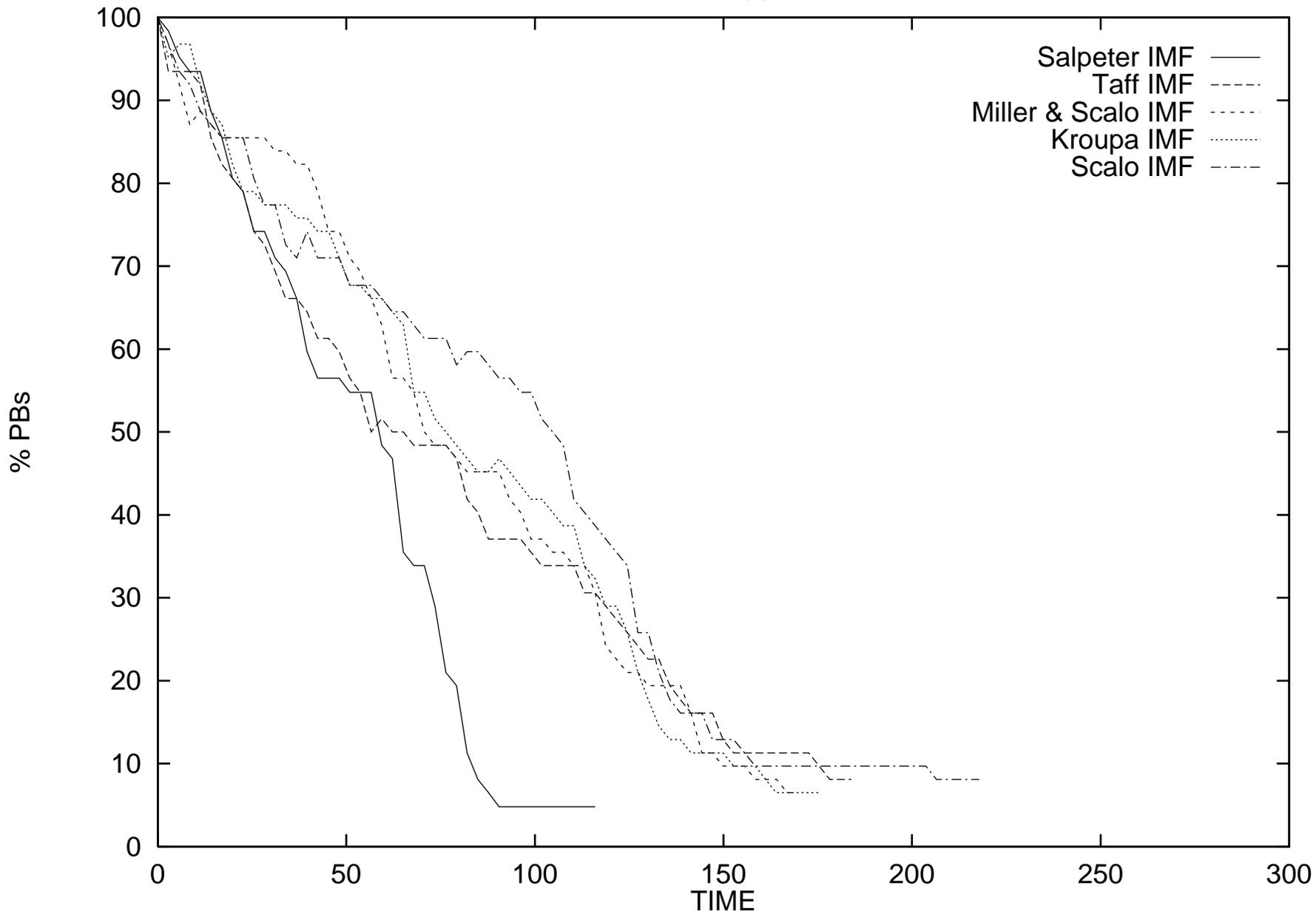
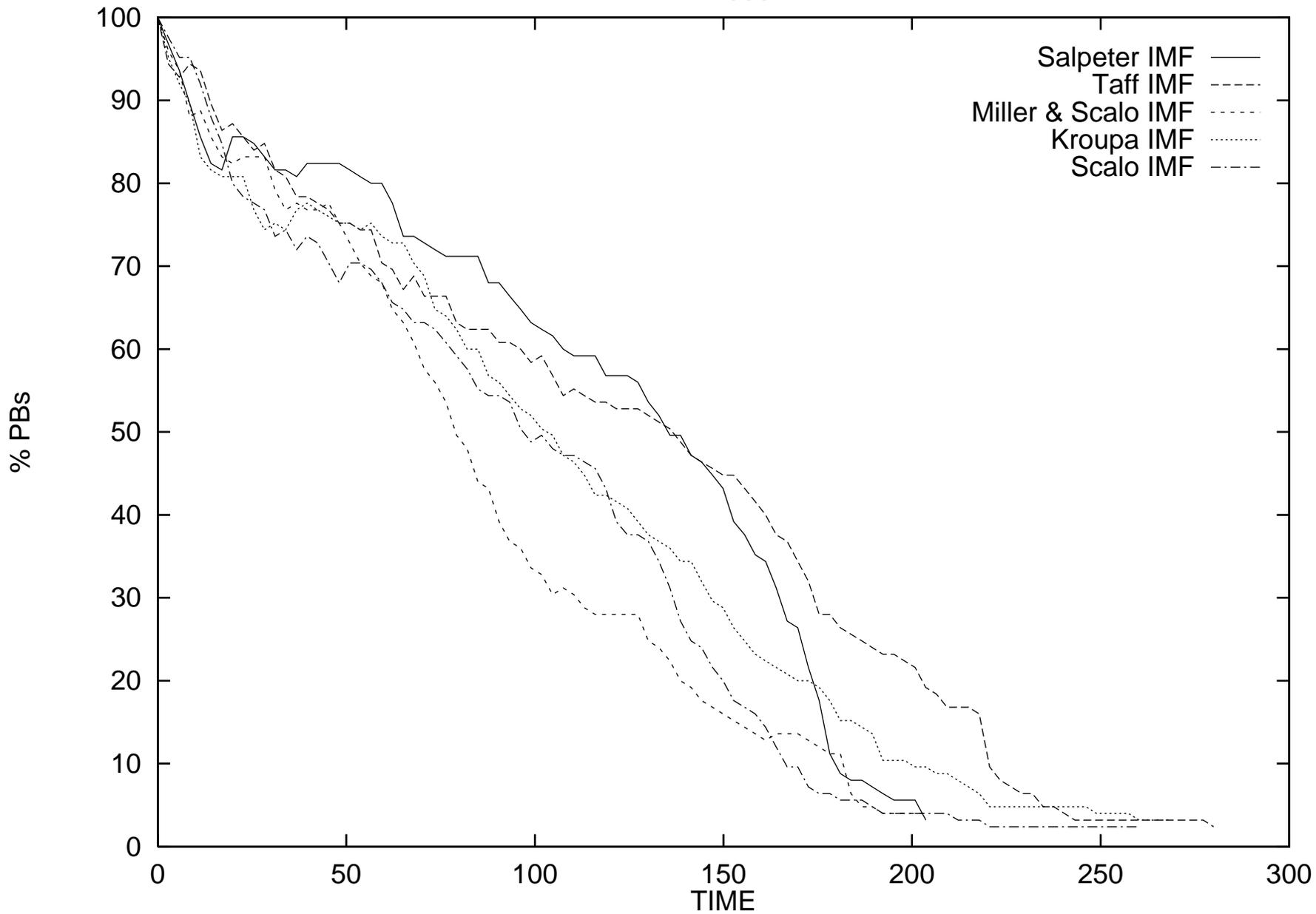
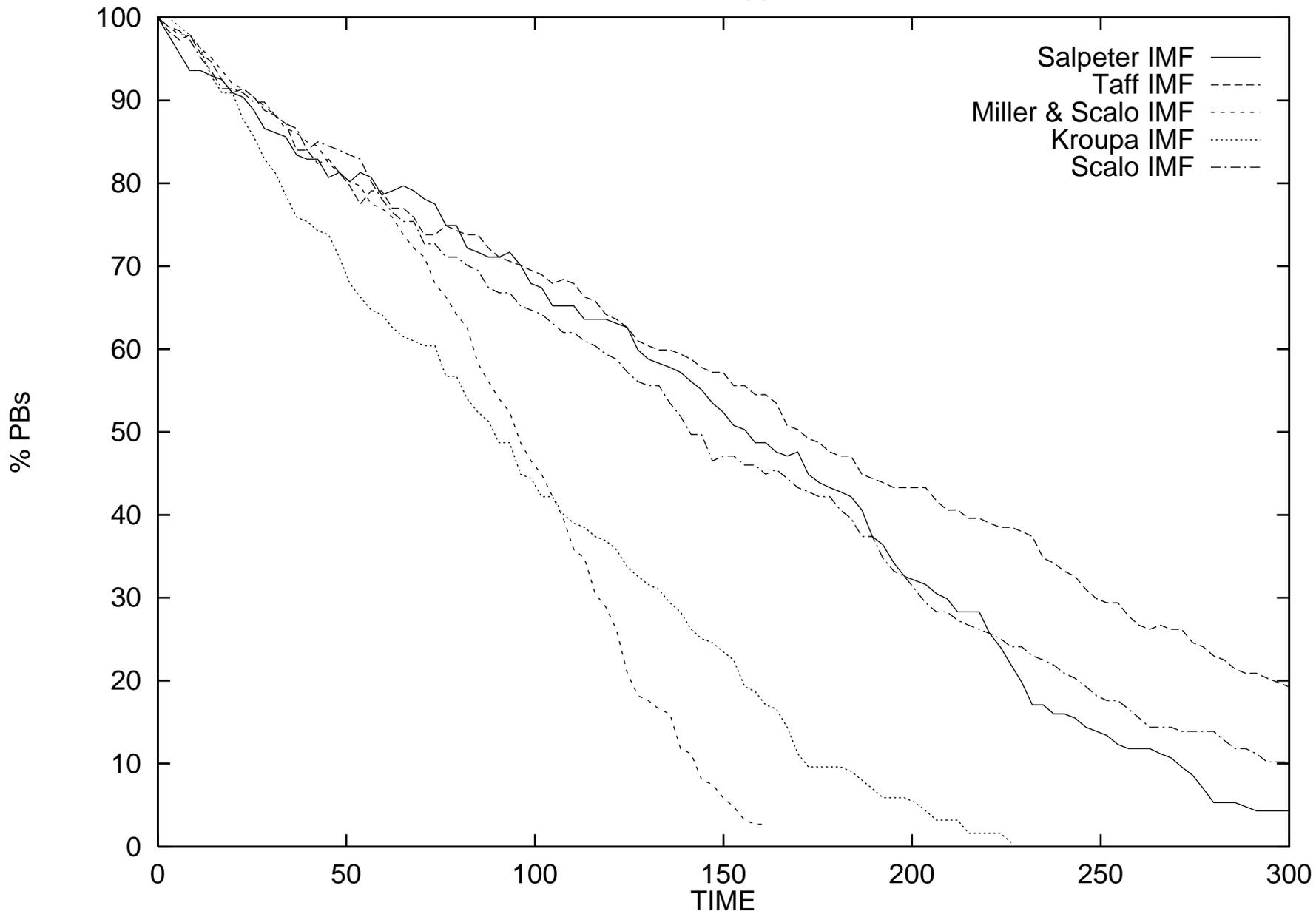
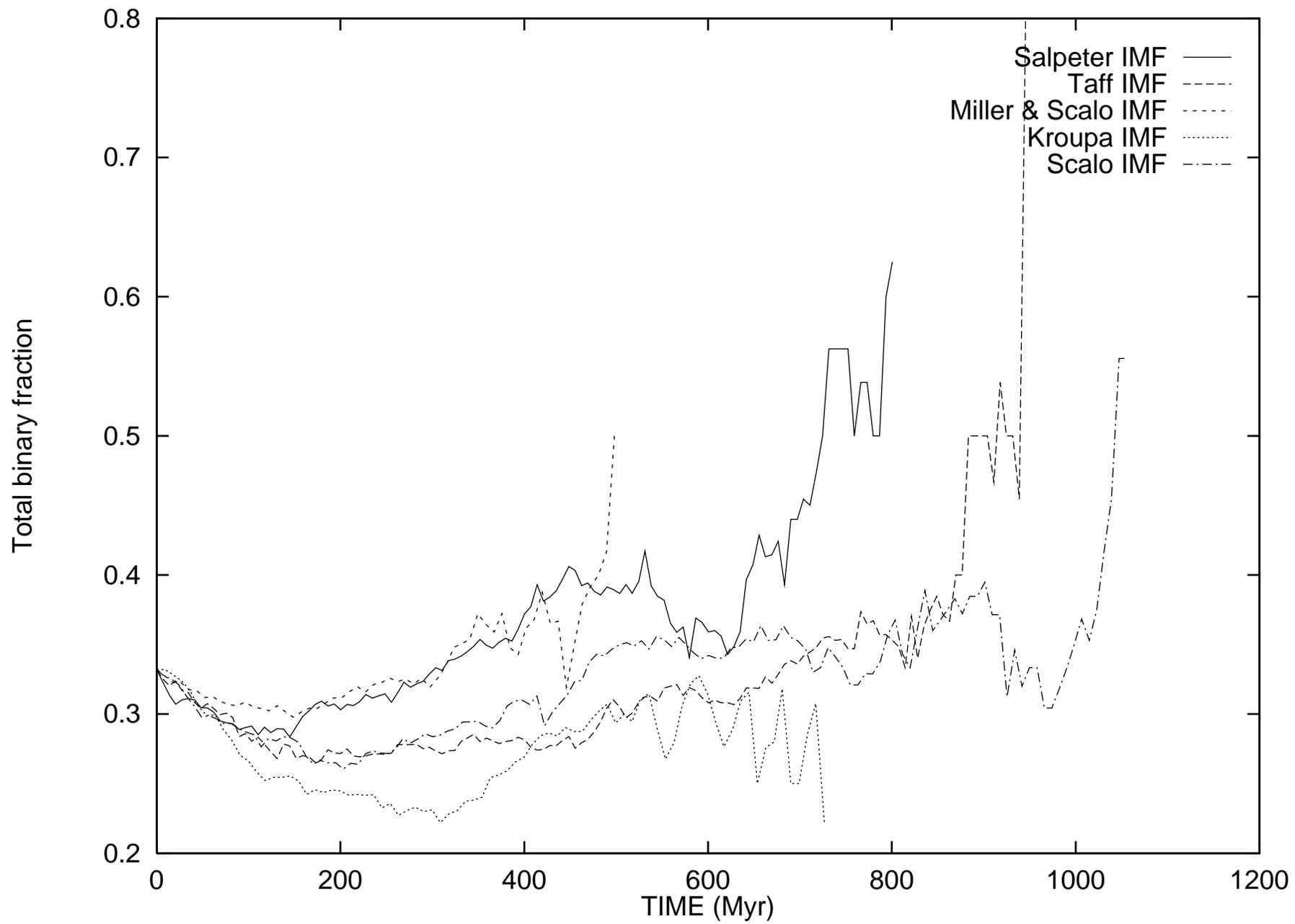
\end{document}